# Holographic position uncertainty and the quantum-classical transition

## C. L. Herzenberg

**Abstract**

Arguments based on general principles of quantum mechanics have suggested that a minimum length associated with Planck-scale unification may in the context of the holographic principle entail a new kind of observable uncertainty in the transverse position of macroscopically separated objects. Here, we address potential implications of such a position uncertainty for establishing an additional threshold between quantum and classical behavior.

**Introduction**

Recent studies have combined the holographic principle with the concept of spacetime as changing character and becoming spatially discrete or pixelated at the Planck level, leading to the introduction of a new type of uncertainty which is currently being examined in experiments.[1-3]

At distances and time intervals larger than the Planck scale, spacetime is generally treated as having a smooth and continuous structure, as described in classical physics and special and general relativity. However, the nature of reality at and below the Planck scale appears to differ, although it is at present still not well understood. At these extremely small distances, the structure of spacetime is often regarded as losing its well-defined continuous character and becoming grainy at distances near the Planck length, and as exhibiting highly energetic and noisy behavior at smaller scales. Because of the departures from classical behavior, the fundamental limit for the shortest possible wavelength for electromagnetic waves is usually considered to be comparable to the Planck length.[4] The Planck length provides the length scale at which (in at least some approaches to quantum gravity) the structure of spacetime becomes dominated by quantum effects.[5] At this scale, the concepts of size and distance break down, as quantum indeterminacy seems to become substantially absolute.[4] Thus, the presence of uncertainties in position measurement and other measurements can be expected at distances comparable to or smaller than the Planck length.

The holographic principle holds that a surface enclosing a volume of space encodes all of the information contained in that volume, and thus prescribes that a volume of space can be fully described by what happens on its boundary.[6] However, if the information encoded on a surface is located on Planck length sized areas, it would seem that the amount of information encoded on the surface could be equivalent to the amount of information contained within the volume only if the corresponding regions in the volume become enlarged in comparison with those on the surface. Effectively, uncertainties encoded on the surface would become magnified within the enclosed volume. Thus,

Planck-scale uncertainties on an encoded surface would correspond to larger regions of uncertainty within a holographic image volume.

While this enlargement of regions of uncertainty is analyzed in terms of a holographic uncertainty principle, somewhat similar conclusions can be arrived at from fairly straightforward considerations of wave optics, provided that uncertainty is present at distances below a limiting wavelength.[7,8]

This general approach can be summarized somewhat more specifically as follows: Arguments based on general principles of quantum mechanics have suggested that a minimum length associated with Planck-scale unification may entail a new kind of observable uncertainty in the transverse position of macroscopically separated bodies.[3] In this context, the holographic uncertainty principle implies an irreducible uncertainty of observable states of spacetime itself, and thus of observable nonlocal relative spatial positions in all systems, in directions transverse to macroscopic spatial separation vectors.[1] Furthermore, this uncertainty and the associated possibility of indirect observation of Planck-scale phenomena has lead to experiments to examine holographic noise in interferometers.[1-3]

Here, we address the question of whether such a position uncertainty might have a role in the quantum-classical transition, and, if so, what implications may follow in terms of establishing an additional threshold between quantum and classical behavior.

**Holographic position uncertainty**

The several earlier studies that have examined quantum gravitational uncertainty of transverse position, the measurement of quantum fluctuations in geometry, and the associated holographic noise in interferometers imply the presence of an irreducible uncertainty of observable relative spatial positions in directions transverse to macroscopic spatial separation vectors.[1] The associated uncertainty or blurring is entirely in angle or transverse position, and is not present in radial distances between events with null separation. However, when positions are measured in directions transverse to such a separation, that transverse positional uncertainty should be revealed. This uncertainty would seem to define a resolution limit in 3-space.[1]

More quantitatively, these studies of holographic uncertainty indicate that the transverse positions of objects subject to this effect will exhibit random variations of magnitude about equal to the geometric mean of the Planck scale and the separation of the objects.[1-3] This position uncertainty is given explicitly in terms of the variance by the following equation:[3]

$$\Delta x_L^2 = \lambda_0 L/2\pi \qquad (1)$$

Here, $\Delta x_L$ is the position uncertainty, $\lambda_0$ is the wavelength associated with the frequency cutoff, and L is the separation of the objects.[3] The wavelength associated with the frequency cutoff is taken to be the minimum length associated with Planck-scale unification, which is the Planck length, $\ell_P = (hG/2\pi c^3)^{1/2}$. Here, G is the gravitational constant, h is Planck's constant, and c is the velocity of light.

If we introduce the Planck length $\ell_P$ into Eqn. (1), we find for the position uncertainty:

$$\Delta x_L = (\ell_P L/2\pi)^{1/2} \qquad (2)$$

As noted earlier, it can be seen that the position uncertainty is given approximately (to within a small numerical factor) by the geometric mean of the Planck length and the separation L of the objects. Because the Planck length is such an extremely small distance, this transverse position uncertainty would necessarily be quite small except for cases of extremely large values of the separation distance L.

**Holographic position uncertainty for large distances**

While analyses of holographic noise have examined the transverse uncertainty in the context of two objects (such as interferometer mirrors) separated at experimental distances, we will instead look into the possibility that one of the two objects under consideration is very distantly removed from a local object of primary interest. Thus we are concerned primarily with a single local object, while another associated object (or perhaps a multitude of other objects located in various different directions and at various extremely large distances) would be present but very distantly removed from the primary object.

We will consider the secondary object or objects to be located at an extremely remote distance. For estimation purposes, we will take as a separation distance the Hubble radius, which gives a rough measure of the radial extent of the observed universe. Then we will have $L = R_H$, where $R_H$ is the Hubble radius, which can be written in the form $R_H = c/H_0$, where $H_0$ is the Hubble constant. We can then reexpress Eqn. (2) in these terms, and find that the position uncertainty for an object with respect to another object or objects located remotely at a distance comparable to the Hubble radius would be given by:

$$\Delta x \approx (c\ell_P/2\pi H_0)^{1/2} \qquad (3)$$

In the case of multiple secondary objects, averaging with respect to direction and distance would be required in order to obtain an improved evaluation of position uncertainty.

It can be seen that the quantities on the right hand side of Eqn. (3) are all determinate. Therefore, we can evaluate an explicit numerical value for this position uncertainty. Inserting a value for the Planck length of $1.62 \times 10^{-35}$ meters, a value for the speed of light of $3 \times 10^8$ meters per second and a value for the Hubble constant of $2.29 \times 10^{-18}$ per second, we find for the positional uncertainty:

$$\Delta x \approx 1.8 \times 10^{-5} \text{ meters} \qquad (4)$$

That is, the magnitude of this new uncertainty amounts to approximately 0.02 mm, or 20 microns. This should give us an initial rough estimate for an uncertainty that might seemingly be expected to be present for any object, in association with its interactions with very distant objects in the far reaches of the universe.

How might such a new uncertainty be interpreted?

Such an uncertainty in the position of an object might not be apparent with respect to nearby objects (since the values of L associated with these objects would be small), but only in relation to objects in the distant universe.

Since most of the universe is distant from us, and since in classical mechanics, position is generally treated as a parameter valid with respect to the entire universe, it would seem that such an uncertainty would have some interpretable significance in the context of semiclassical physics, and that consideration should be given to attempting to understand the possible implications of any such uncertainty.

**Holographic position uncertainty and the quantum-classical boundary**

What connection, if any, might this new uncertainty have with respect to the question of classical versus quantum behavior?

A simple criterion that can be helpful for considering the quantum or classical behavior of extended objects is the following: an extended object may be regarded as behaving in a largely quantum manner when the uncertainty in its location extends well beyond its boundaries; whereas an object can be regarded as behaving in a largely classical manner when the uncertainty in its location is much smaller than its actual size.[9-13]

Let us consider applying this criterion to an extended object having a size characterized approximately by a linear extent $L_E$. Then this criterion for a boundary separating quantum from classical behavior for an extended object having a position uncertainty $\Delta x$ may be expressed as:

$$L_E \approx \Delta x \qquad (5)$$

Thus, any extended object having a size larger than $L_E$ would be expected to behave in a classical manner with respect to the phenomenon under consideration, since the object's size would exceed the magnitude of the uncertainty due to the effect, while any extended object having a size smaller than $L_E$ might be expected to exhibit quantum behavior unless brought into classicality by another effect, such as quantum decoherence.[14]

Using the criterion for a quantum-classical boundary for an extended object expressed in Eqn. (5), and inserting the holographic uncertainty given in Eqn. (3), we find that a quantum-classical boundary associated with this holographic uncertainty for an extended object would be expected to appear at a critical size $L_{Ec}$ such that:

$$L_{Ec} \approx (c\ell_P/2\pi H_0)^{1/2} \qquad (6)$$

Furthermore, using the numerical result given in Eqn. (4), we might surmise that extended objects smaller than about 20 microns might be expected to exhibit quantum behavior associated with this holographic uncertainty, whereas objects having sizes exceeding about 20 microns could be expected to have classical rather than quantum mechanical behavior with respect to the holographic position uncertainty of such an object in relationship to other objects in the distant universe.

**Further discussion of the quantum-classical boundary**

Thus, to the extent that the discussion in this paper is valid, this holographic position uncertainty would seem to set an approximate bound on size such that objects smaller than roughly 20 microns would be expected to exhibit quantum uncertainty due to holographic effects, except to the extent that they are affected by other phenomena that can bring about classical behavior, such as quantum decoherence.[14]

Other quantum-classical thresholds originating from or relating to large-scale properties of the universe have been proposed in the past; these include a boundary based on the information content of the universe and boundaries based on effects associated with the expansion and the limited duration of the universe.[9-13,15,16]

Davies has proposed that classicality may emerge above a certain level of complexity rather than mass or size.[15] He suggests that classicality can emerge in sufficiently complex quantum systems; and specifically in a system requiring more bits to specify it than the upper bound on the information content of our universe.[15]

The upper bound on the information content of the universe according to the holographic principle can be evaluated as ¼ the ratio of the surface area of the two-dimensional boundary of the universe to the Planck area.[15] This upper bound on the information content of the universe, in accordance with the holographic principle, has been evaluated as $I_{universe} = 10^{122}$ bits.[15] (If we make a crude estimate using for the radius of the observable universe to be approximately $10^{26}$ meters, and taking the Planck area to be about $2 \times 10^{-70}$ meters, that is, equal to the square of the Planck length, we get an estimate in approximate agreement with the value that Davies presents.)

The emergence of classicality would then correspond to systems requiring an information content exceeding $I_{universe} = 10^{122}$ bits, the upper bound on the information content of the universe according to the holographic principle.[15] Thus, Davies suggests that a quantum state with more components than about $n = \log_2 I_{universe}$ will require more bits of information than can be accommodated in the entire observable universe.[15] Accordingly, Davies has interpreted the threshold system as one that corresponds to an entangled state of more than about n = log$_2$ I$_{universe}$ or 400 particles.[15] Thus Davies' threshold would seem to call for quantum behavior for objects of fewer that roughly 400 particles, and classical behavior for larger objects. But for ordinary matter, depending upon how one counts entangled particles, a quantum system of 400 particles might consist of a reasonably small molecule or a nanoparticle having a diameter of somewhat less than 10 atoms.

In other studies, constraints on quantum systems imposed by the expansion or limited temporal duration of the universe have been studied and their possible effects on a quantum-classical threshold have been examined.[9-13] In ordinary matter, these effects appear to lead to a possible quantum-classical threshold for ordinary objects very roughly in the range of sizes of about 0.1 millimeter, or for objects having masses very roughly in the range of a microgram.

As we have noted earlier, various types of local decoherence effects can bring about classical behavior from quantum behavior.[14] It now would seem that there may also be several built-in effects associated with the structure of the universe itself that may provide for quantum behavior for smaller objects while leading to classical behavior for objects in the range of sizes between small molecules and ordinary macroscopic objects.

**Summary and conclusions**

This new transverse holographic uncertainty would seem by itself to cause quantum mechanical behavior for objects smaller than about 20 microns (as a very rough estimate), while larger objects would not be so constrained by this effect and could behave in a classical manner.